\documentclass[a4paper,11pt]{article}

\usepackage[english]{babel}
\usepackage{jheppub}

\usepackage{graphicx}
\usepackage{cancel}
\usepackage{MnSymbol}
\usepackage{bbold}
\usepackage{amsmath,amssymb,amsfonts}
\usepackage{slashed}
\usepackage{caption}
\usepackage{appendix}
\usepackage{hyperref} 
\usepackage{multirow}
\hypersetup{colorlinks=true, linkcolor=black, urlcolor=blue}
\usepackage{comment}
\usepackage{float}
\usepackage{mathtools}

\usepackage{amstext} 
\usepackage{array}   
\newcolumntype{L}{>{$}l<{$}} 

\usepackage{booktabs}

\usepackage{geometry}
\usepackage{color}
\definecolor{gray}{rgb}{0.5,0.5,0.5}

\newcommand\lsim{\mathrel{\rlap{\lower4pt\hbox{\hskip1pt$\sim$}}
    \raise1pt\hbox{$<$}}}
\newcommand\gsim{\mathrel{\rlap{\lower4pt\hbox{\hskip1pt$\sim$}}
    \raise1pt\hbox{$>$}}}

\newcommand{\beq}{\begin{equation}}
\newcommand{\eeq}{\end{equation}}
\newcommand{\bea}{\begin{eqnarray}}
\newcommand{\eea}{\end{eqnarray}}
\newcommand{\bem}{\begin{pmatrix}}
\newcommand{\eem}{\end{pmatrix}}

\newcommand{\bet}{\begin{itemize}}
\newcommand{\eet}{\end{itemize}}
\newcommand{\ben}{\begin{enumerate}}
\newcommand{\een}{\end{enumerate}}

\definecolor{myGreen}{rgb}{0.39, 0.65, 0.46}
\definecolor{myBlue}{rgb}{0.17, 0.26, 0.65}
\definecolor{myRed}{rgb}{0.85, 0.0, 0.0}

\makeatletter
\gdef\@fpheader{}
\makeatother

\begin{document}

\title{ Parity Solution to  the Strong CP Problem
and a Unified Framework for Inflation, Baryogenesis, and Dark Matter}

\author[a]{K.S.~Babu,}
\author[b]{Rabindra N.~Mohapatra,}
\author[c]{Nobuchika Okada}

\affiliation[a]{Department of Physics, Oklahoma State University, Stillwater, OK, 74078, USA}
\affiliation[b]{Maryland Center for Fundamental Physics and Department of Physics, University of Maryland, \\College Park, MD 20742, USA}
\affiliation[c]{Department of Physics and Astronomy, University of Alabama, Tuscaloosa, AL 35487, USA}

\emailAdd{babu@okstate.edu}
\emailAdd{rmohapat@umd.edu}
\emailAdd{okadan@ua.edu}

\abstract{
It has been known for some time that asymptotic parity invariance of weak interactions can provide a solution to the strong CP problem without the need for the axion.  Left-right symmetric theories which employ a minimal Higgs sector consisting of a left-handed and a right-handed doublet is an example of such a theory wherein all fermion masses arise through a  generalized seesaw mechanism. In this paper we present a way to understand the origin of matter-antimatter asymmetry as well as the dark matter content of the universe in these theories using the Affleck-Dine (AD) leptogenesis mechanism and inflaton decay, respectively. Three gauge singlet fermions are needed for this purpose, two of which help to implement the Dirac seesaw for neutrino masses while the third one becomes the non-thermal warm dark matter candidate. A soft lepton number breaking term involving the AD scalar field is used to generate lepton asymmetry which suffers no wash-out effects and maintains the Dirac nature of neutrinos. This framework thus provides a unified description of many of the unresolved puzzles of the standard model that require new physics.
}

\maketitle

\section{Introduction}
The Standard Model, while highly successful in confronting a variety of experimental data, needs extension to understand several deficiencies. These include a lack of understanding of small neutrino masses, a dark matter candidate with the right properties, as well as the origin of matter-antimatter asymmetry of the universe, all of which require new ingredients. On the theoretical side, quantum chromodynamics admits an operator that violates Parity ($P$) and Charge-Parity ($CP$) whose strength is characterized by an arbitrary dimensionless parameter, $\overline{\theta}$. In presence of this parameter the neutron would acquire an electric dipole moment (nEDM) $d_n\sim 10^{-16} ~\overline{\theta}$ e-cm. Current experimental bounds on nEDM imply that $\overline{\theta}$ must be a very small number, $\overline{\theta} \leq 10^{-10}$. A lack of understanding of the smallness of this parameter, which could have been of order unity,  is the strong $CP$ problem. The most popular solution to this problem is the Peccei-Quinn (PQ) proposal \cite{PQ} which postulates a global axial symmetry acting on the quark fields and predicts a near massless particle, the axion \cite{W,W1}, which is the pseudo-Nambu-Goldstone boson associated with the spontaneous breaking of the PQ symmetry.
However, there is so far no evidence for the axion despite many dedicated experimental searches, which should of course continue to fully test this hypothesis. The popular invisible axion models \cite{kim,svz,dfs,z} which evade all laboratory constraints seem to have a ``quality problem" \cite{K, holman, barr}, since non-perturbative gravitational effects which are believed to violate all global symmetries including the PQ symmetry would induce an unacceptably large $\overline{\theta}$, thus destabilizing the axion solution. Evading this would require a fine-tuning at the level of $10^{-50}$ for these Planck-suppressed operators, assuming naive scaling of these operators with the inverse Planck scale.  In view of these issues, it is not premature to explore alternatives 
to the axion solution to the strong $CP$ problem.

A ``no axion" solution to the strong CP problem  was proposed shortly after the axion solution was suggested~\cite{BT, MS1}. It is based on the idea that the ultimate theory of weak interactions may be parity symmetric and the observed parity violating weak interaction may be a long distance effect resulting from the spontaneous breakdown of parity. 
These theories are the so-called left-right (LR) symmetric theories of weak interactions \cite{LR1,LR2,LR3}. In this class of theories the QCD contribution to $\overline{\theta}$, denoted as  $\theta_{QCD}$, vanishes owing to parity invariance. Secondly, the flavor contribution to $\overline{\theta}$, given as $\theta_{QFD} \equiv {\rm Arg} ({\rm Det}[M_uM_d])$ could also be vanishing, since $P$-invariance makes the Yukawa coupling matrix of the quarks to be hermitian.  This, however, would require the vacuum expectation value (VEV) of the scalar fields coupling to the quarks to be real, so that the quark mass matrix is also hermitian.  If the Higgs potential admits such a vecuum structure, then $\overline{\theta} = 0$ at tree level.  The next issue is to check if the spontaneous $P$-breaking 
induces a nonzero $\overline{\theta}$ that is in the acceptable range, and if it satisfies this demand, one has a solution to the strong CP problem without the need for the axion.  It, however, turns out that with the minimal fermion content of LR theories, the Higgs bidoublet needed for fermion mass generation  $\Phi~({\bf 2},{\bf 2},0)$ under $SU(2)_L \times SU(2)_R \times U(1)_{B-L}$, would acquire a complex VEV, which spoils the promise of such models to solve the strong CP problem. This hurdle, however, can be overcome by supplementing LR symmetry with certain exterior symmetries such as supersymmetry \cite{MR1,MR2,MR3,BDM}. A natural question then is whether there are models wherein parity symmetry alone can solve the strong $CP$ problem. 
The answer to this question was provided in the affirmative in Ref.~\cite{BM}, where a minimal Higgs sector of the LR models involving a left-handed and a right-handed doublet fields was proposed.  The VEVs of these Higgs fileds can be taken to be real by independent $SU(2)_L$ and $SU(2)_R$ rotations, unlike the case of a Higgs bidoublet field.  Fermion masses arise in this setup via a universal seesaw involving vector-like quarks and leptons \cite{DW}. In this class of  models, no extra symmetry other than parity is needed to have a small and finite $\theta$, which arises only at the two-loop level \cite{BM}. Detailed computation of the induced $\overline{\theta}$ has been carried out and shown to be in the acceptable range in Ref. \cite{hisano} recently. Furthermore, satisfying the ``quality" constraint in this model requires that the parity breaking scale must be less than about $(100-1000)$ TeV \cite{Berezhiani:1992pq}, which puts associated physics in the observable range of planned experiments~\cite{craig}. Other aspects of this model such as possible grand unification~\cite{HH},  dark matter and leptogenesis \cite{DHH}, flavor constraints \cite{dcruz}, relevance to $W$-boson mass shift and the CKM unitarity  \cite{Babu:2022jbn} and baryogenesis \cite{hariw} have also been recently discussed. In this framework the neutrino can naturally be a Dirac fermion with its mass arising as two-loop radiative corrections \cite{Babu:1988yq,babu1}. It is the class of these models that we focus on in this paper.

The goal here is to study whether, in these models, one can understand the origin of matter-anti-matter asymmetry and dark matter of the universe. We find that by extending the model to include three generations of gauge singlet fermions (denoted by $N_a$) and a complex scalar field carrying a nonzero lepton number ($\Phi$), this goal can be achieved. The addition of singlet fermions has a certain appeal since they make the vector-like sector of the model quark-lepton symmetric and they also provide a Dirac seesaw for neutrinos. On the experimental side, the two major predictions of this model are that the neutrinos are Dirac fermions and the dark matter is warm.

The paper is organized as follows: In Sec. 2, we briefly review the model along with the new element that includes the gauge singlet neutrinos $N_a$. In Sec. 3, we show how small Dirac neutrino masses arise in the model via Dirac seesaw;  in Sec. 4, we review the general picture of the evolution of the universe until the inflaton decay and outline possible scenarios for leptogenesis. Sec. 5 is devoted to more details about baryogenesis and dark matter relic density generation. In Sec. 6, we give some comments on the model and conclude in Sec. 7. An appendix summarizes some other scenarios for baryogenesis and dark matter for different regions of parameters of the model.

\section{The model}
The model is based on the left-right gauge group, 
$SU(3)_c \times SU(2)_L\times SU(2)_R\times U(1)$, and has the following fermion assignments: 
\begin{eqnarray}
Q_{L,R}~=~\left(\begin{array}{c}u\\d\end{array}\right)_{L,R};~~~\ell_{L,R}~=~\left(\begin{array}{c}\nu\\e\end{array}\right)_{L,R}~.
\end{eqnarray}
Here $Q_L$ transforms under the gauge group as $({\bf 3,2,1},+\frac{1}{3})$, while $Q_R$ transforms as $({\bf 3,1,2},+\frac{1}{3})$, while the leptons are $\ell_L({\bf 1,2,1},-1)$ + $ \ell_R({\bf 1,1,2},-1)$. 
This is
supplemented by vector-like singlet quarks $U_{L,R}({\bf 3},{\bf 1}, {\bf 1},+\frac{4}{3})$,
$D_{L,R}({\bf 3},{\bf 1},{\bf 1},-\frac{2}{3})$, and 
leptons $E_{L,R}({\bf 1},{\bf 1},{\bf 1},-2)$. These vectro-like fermions are used to generate quark and lepton masses via a generalized seesaw.
  
The Higgs sector of the model consists of just two fields, an $SU(2)_L$ doublet  and an $SU(2)_R$ doublet
denoted by $\chi_L({\bf 1},{\bf 2},{\bf 1},-1)$ and $\chi_R ({\bf 1},{\bf 1},{\bf 2},-1)$.  The Higgs potential involving these fields is given by
\begin{eqnarray}
V(\chi_L, \chi_R)~=~-\mu^2_L\chi^\dagger_L\chi_L -\mu^2_R\chi^\dagger_R\chi_R+\lambda_1 (\chi^\dagger_L\chi_L+\chi^\dagger_R\chi_R)^2 +\lambda_2(\chi^\dagger_L\chi_L)(\chi^\dagger_R\chi_R). 
\end{eqnarray}
Here a soft breaking of parity symmetry is assumed that makes $\mu_L^2 \neq \mu_R^2$.  Such a breaking allows for a vacuum structure where $\langle \chi^0_L \rangle=v_L \ll \langle \chi^0_R \rangle=v_R$ can be realized. Notably, both VEVs can be made real by gauge rotations, which helps in the solution to the strong CP problem. 

The Yukawa Lagrangian that generates the charged fermion masses in the model is given by
\begin{eqnarray}
{\cal L}_Y~&=&~y_u\,\overline{Q}_L\chi_L U_R+y_u\,\overline{Q}_R\chi_R U_L~+~y_d\,\overline{Q}_L\tilde{\chi}_L D_R+y_d\,\overline{Q}_R\tilde{\chi}_R D_L~\\\nonumber
&&+~y_e\,\overline{\ell}_L\tilde{\chi}_L E_R+y_e\,\overline{\ell}_R\tilde{\chi}_RE_L+ M_U \overline{U}_L U_R + M_D \overline{D}_L D_R + M_E \overline{E}_L E_R + h.c. 
\end{eqnarray}
Here $\tilde{\chi}_{L,R} = i \tau_2 \chi_{L,R}^*$.  
The resulting mass matrices for the up-type quarks, down-type quarks and charged leptons are given by
\begin{eqnarray}
{\cal M}_{f} = \left( \begin{matrix}  0 & y_{f} \,v_L \cr y^\dagger_{f} \,v_R & M_{F} \end{matrix}  \right),~~~f=(u,d,e); ~~~F = (U,D,E)~.
\label{eq:Mf}
\end{eqnarray}
Owing to parity symmetry $M_F = M_F^\dagger$ in this setup. 
It is clear from Eq. (\ref{eq:Mf}) that the determinant of the quark mass matrix is real and therefore $\overline{\theta}$ vanishes at the tree level. It was shown in Ref. \cite{BM} that in this model finite and small $\overline{\theta}$ arises only through two-loop diagrams. The model thus
provides a pure parity solution to the strong CP problem without any extra ingredients unlike the left-right models with Higgs bi-doublet fields.

In order to solve the neutrino mass problem, we extend the model by adding gauge singlet neutral leptons $N_{L,R}:({\bf 1},{\bf 1},{\bf 1},0)$ per generation. 
In this case, the neutrinos can either be Dirac or Majorana type. We choose the Dirac alternative by requiring that the model have a lepton number symmetry. It is interesting that the lepton number is an anomaly-free 
gauge symmetry and therefore is free from Planck scale corrections that break it. To see this we note that $B-L$ symmetry has no gauge anomalies in this setup.  The $B-L$ charge of all quarks, including the vector-like quarks, is $1/3$ while those for leptons, including the vector-like lepton is $-1$. These charges are distinct from the gauged $U(1)$ charges of the fermions. Such an assignment has no mixed anomaly with the gauged $U(1)$ of the theory.  A $Z_4$ subgroup of this $B-L$ symmetry will remain unbroken even after soft breaking via dimension-two terms in the scalar potential involving the inflaton field (see later).  This $Z_4$ symmetry will guarantee that neutrinos will be strictly Dirac fermions in the framework. Also, as will be discussed below, we require two of the $N$s to couple to SM fields while the third one fully decouples from SM fields, so that it can play the role of dark matter.

The Yukawa Lagrangian for the neutral fields of the model is given by
\begin{eqnarray}
{\cal L}_Y^\nu=
y_\nu(\,\overline{\ell}_L\chi_LN_R+\overline{\ell}_R\chi_R N_L)
+ M_N\overline{N}_L N_R+ ~h.c. , 
\end{eqnarray}
where we have assumed a lepton number symmetry. As stated, this symmetry is a discrete gauge symmetry and is hence protected from Planck scale corrections.
The neutrinos in this case are Dirac particles.

\section{Neutrino mass} 
In universal seesaw models, neutrino masses can arise from various mechanisms. In the minimal model with only vector-like fermions  $(U, D, E)$,  and no $N$ fields, there are two-loop diagrams which give rise to neutrino Dirac masses~\cite{Babu:1988yq,babu1}. Since this scenario has exact lepton number symmetry, the neutrinos in this case are Dirac fermions. A second source of neutrino mass arises when a gauge singlet fermion $N_{L,R}$ is added to the model, as we do in the present work. In this case, the neutrino masses can either be Majorana or Dirac type.
The neutrino mass matrix in the general universal seesaw framework is given in Ref. ~\cite{DW,yongchao,dcruz}. 
Here, however, we are interested in the Dirac neutrino possibility; so we choose the following $4\times 4$ block mass matrix for neutrinos and SM singlet fermions: 
\begin{eqnarray}
M_\nu~=~\left(\begin{array}{cccc}\overline{\nu_L}  &  \overline{N_L} & 
\overline{\nu_R^{~c}} & \overline{N_R^{~c}} \end{array}\right) \left(\begin{array}{cccc} 0 & 0 & 0 & y_\nu v_L\\0 & 0 & y_\nu v_R & M_N\\
0 & y_\nu v_R & 0 & 0 \\ y_\nu v_L & M_N& 0 & 0 \end{array}\right)\left(\begin{array}{cccc}\nu_L^{~c} \\ N_L^{~c}\\ \nu_R \\ N_R\end{array}\right) ,
\end{eqnarray} 
where  $y_\nu$ and $ M$ are $3\times 3$ matrices.
Here, both the left and the right chirality neutrinos combine to form the light Dirac neutrinos with their mass given by Dirac seesaw form (see ~\cite{dirac1, dirac2, dirac3, dirac4, dirac41, dirac43, dirac42, dirac5, dirac6, dirac7} for a small sample of the vast literature on Dirac seesaw):
\begin{eqnarray} 
M_\nu~=~ \frac{y^2_\nu v_Lv_R}{M_N}.
\end{eqnarray}
Although the charged fermion masses arising from Eq. (\ref{eq:Mf}) also have a similar seesaw form, one can understand the relative smallness of neutrino masses by choosing $y_\nu \sim 10^{-4} \,y_e$, or alternatively by choosing $M_N \gg M_E$, or a combination of the two.

We show below how the origin of matter as well as dark matter, can be explained in the model with the Dirac seesaw. For this purpose, we will let only two heavier singlet fermions $N_{2,3}$ couple to the SM sector and keep the lightest one ($N_1$) decoupled from the SM. This will have the consequence that one of the light neutrinos will be massless in both left and right handed helicity. This will lead to interesting consequences for cosmology that we discuss below.

\section{Evolution of the universe}
Before proceeding further, we first review the profile of the evolution of the universe in our model. A key feature of this discussion involves the complex singlet scalar $\Phi$ that we add to the model. This will help us to implement inflation, generate dark matter density and explain the origin of matter via Affleck-Dine leptogenesis \cite{AD}.
We will assume that the field $\Phi$ carries lepton number $L=-2$. We then add the following terms to the Lagrangian: 
\begin{eqnarray}
{\cal L}_\Phi'~=~\sum_{a=1}^3 f_{N,a}\Phi(N_L^a N_L^a + N_R^a N_R^a ) + h.c. 
-M^2_\Phi|\Phi|^2 -\lambda_\Phi |\Phi|^4 -\epsilon M^2_\Phi (\Phi\Phi+h.c.). 
\label{Eq:NN_coupling}
\end{eqnarray}
Here the $\epsilon$ term breaks lepton number sofly by 4 units, leving a $Z_4$ subgroup of $L$ intact. 
We call  $\Phi$ as the AD field, which also plays the role of inflaton. 
The dynamics of inflation arises from the non-minimal gravity coupling of the $\Phi$ field given  \cite{berzukov}:
\begin{eqnarray}
{\cal L}_{g}~=~-\frac{1}{2}\int d^4 x  \,\sqrt{-g} \, \left[M^2_P+ 2 \xi |\Phi|^2 \right] R ,
\end{eqnarray}
where $M_P=2.4 \times 10^{18}$ GeV is the reduced Planck mass. 
This is in the Jordan frame, and making a Weyl transformation 
via $g^J_{\mu\nu}\to \left(1+ 2 \frac{|\Phi|^2 }{M^2_P} \right)g^J_{\mu\nu}\equiv g^E_{\mu\nu}$, 
we can go to the Einstein frame, where the potential for $\Phi$ becomes 
\begin{eqnarray}
V^E(\Phi)~=~\frac{V^J(\Phi)}{\left(1+2 \xi\frac{|\Phi|^2}{M^2_P}\right)^2}.
\end{eqnarray}
As we see from the shape of the potential in the above equation, it is flat for $\Phi > M_P/\sqrt{\xi}$, and this then provides a model for inflation, 
leading to an exponential expansion of the universe. 
As inflation proceeds, the value of $\Phi$ becomes smaller as it rolls down the potential, 
and inflation ends when $\Phi$ becomes lower than $M_P/\sqrt{\xi}$.

We implement Affleck-Dine leptogenesis in this model by using the Lagrangian as above \cite{Inf1,Inf2}.
Following the particular implementation of the AD mechanism as in Ref. \cite{gorbunov, stubbs, MO1}, we first generate the asymmetry in lepton number carried in the $\Phi$ field, followed by its decay,
$\Phi\to N_L N_L, N_R N_R$. The latter creates an equal asymmetry in the $N_{L,R}$-number. 
When $N_{L,R}$ decay to SM fermions due to the Yukawa couplings in Eq.~(\ref{Eq:NN_coupling}), 
this lepton asymmetry is transferred to the SM fermions and is later converted to baryon number by the sphalerons.
 
 The various stages in the evolution of $\Phi$ in the universe leading to leptogenesis are as follows:
 
 \begin{enumerate} 
 
\item  In the very early universe when $|\Phi| \gtrsim M_P/\sqrt{\xi}$, the non-minimal coupling in the Einstein frame leads to a constant $V^E(\Phi)$, and drives inflation, as stated above.

\item In the second phase, as the field  $|\Phi|$ has rolled sufficiently down the potential to its value less than $M_P/\sqrt{\xi}$, the effect of the non-minimal coupling becomes unimportant and inflation ends. 
The value of $|\Phi|$ is still large 
and the dominant term in the potential driving the evolution of the $|\Phi|$ is the $\lambda |\Phi|^4$ term.  
At the beginning of this stage, the real and imaginary parts of the field are already different, owing to the $\epsilon$ term in Eq. (4.1).
This asymmetry survives the evolution of the $\Phi$ field and eventually leads to the baryon asymmetry of the universe. This is the key idea in AD baryogenesis.

\item As the universe evolves further, $\Phi$ becomes smaller and the third stage begins where the quadratic term in the potential dominates over the quartic term. This leads to an oscillatory behavior of $|\Phi|$ and the universe behaves like it is matter dominated. This approximation of transition of the potential from being quartic dominated to quadratic dominated is called the threshold approximation in \cite{stubbs}.

\item The fourth  stage is when the AD field decays to $N_{1,2,3}N_{1,2,3}$ states and in our scenario, the $N_{2,3}$-fields decay immediately to standard model particles. The $N_1$ is assumed to have no interaction with the SM fields. As a result, it remains stable after production and becomes the dark matter of the universe. We describe the details of how this works out in the two subsequent sections. 
At this stage, the $\Phi$ asymmetry gets transferred to $N$ asymmetry. In the formulation given in Ref.~\cite{stubbs, MO1}, the decay products are assumed to quickly thermalize, and the SM baryon asymmetry is generated due to sphaleron interactions.

\end{enumerate}

In the following two sections, we proceed to discuss the details of how matter-anti-matter asymmetry is generated in our model via the Affleck-Dine mechanism and how one obtains the dark matter relic density.
As noted above, for both of these purposes, we are using the singlet fermions $N_{1,2,3}$, with the heavier ones $N_{2,3}$ being responsible for leptogenesis whereas the lightest one $N_1$ becomes the dark matter of the model. As noted above, $N_{2,3}$ couple to the SM fields, but $N_1$ does not. 
As given in Eq.~(\ref{Eq:NN_coupling}), the $\Phi$ field coupling to all three of them with arbitrary coupling values has the form: 
\begin{eqnarray}
{\cal L}_N~=~\sum_a f_{N,a}\Phi N_aN_a+ h.c.,
\end{eqnarray}
where we have chosen a basis in which the $f$ matrix is diagonal, and we have suppressed the $L,R$ indices. Clearly, $\Phi$ carries lepton number $L=-2$.

\subsection{Possible scenarios for baryogenesis and dark matter} 
The key parameters of the model in discussing baryogenesis are the masses $M_\Phi$ and $M_{N_{2,3}}$; the decay widths of $\Phi$ and $N_{2,3}$ i.e.~$ \Gamma_\Phi$ and $\Gamma_{N_{2,3}}$, 
and the $\Phi NN$ coupling $f_N$ which determines whether the $N$ fields resulting from the decay of $\Phi$ are in thermal equilibrium or not. There are eight possibilities depending on the relative magnitudes of the above parameters. They can be classified as follows:
\begin{eqnarray}
&(1)& \Gamma_\Phi > \Gamma_N , \\\nonumber
&(2)& \Gamma_\Phi < \Gamma_N. 
 \end{eqnarray}
Using the seesaw formula, $M_\nu \sim  5 \times 10^{-11}$ GeV and $v_R =10^6 $ GeV, which we choose as our benchmark point, we can write
$\Gamma_N \sim \frac{1}{4 \pi}  \frac{M^2_N}{M_P }$ and   
$\Gamma_\Phi \sim  \frac{1}{4 \pi}  f^2_N M_\Phi$. The above equation can then be rewritten as   two separate  possibilities:
\begin{eqnarray}
 &(1)&f^2_N > \frac{M^2_N}{M_\Phi M_P}~~ ({\rm delayed~decay~of}~N), \\\nonumber 
 &(2)& f^2_N < \frac{M^2_N}{M_\Phi M_P}~~({\rm immediate~decay~of}~N).
  \end{eqnarray}  

Next, we consider two cases: depending on whether $N$s produced are in thermal equilibrium (case  (i)) or not  (case (ii)). To check it, we note that the only interactions that the decay products of $\Phi$
i.e.~$N_{1,2,3}$ have at this stage are via the $\Phi$ exchange. It is easy to see that these interactions undergo resonance enhancement. So, to bring the $N$s into thermal equilibrium, there have to be restrictions on $f_N$. To quantify this, consider the $s$-channel contribution to the process $N_iN_i\to N_jN_j$ via $\Phi$ exchange. The cross section for this channel can be written in a Breit-Wigner form as
\begin{eqnarray}
\sigma v
\simeq \frac{f^4_N}{4\pi}\frac{s}{(s-M^2_\Phi)^2+\Gamma^2_\Phi M^2_\Phi}\sim \frac{f^4_N}{4\pi}\frac{1}{\Gamma^2_\Phi},
\end{eqnarray}
since $s \simeq M_\Phi^2$ for the produced $N$s.
To check for equilibrium, we need to calculate $n_N\sigma v$, where $n_N$ is the number density of the decay products $N_{1,2,3}$. We can estimate $n_N\sim 2 \frac{\rho_\Phi}{M_\Phi}\sim \frac{6\Gamma^2_\Phi M^2_P}{ M_\Phi}$ and compare it with the Hubble parameter $H\sim \sqrt{\frac{\rho_\Phi}{3M^2_P}}\sim \Gamma_\Phi$. Using $\Gamma_\Phi\simeq \frac{f^2_N M_\Phi}{4\pi}$, we find that if  $f_N <\frac{M_\Phi}{M_P}$, the $N$'s are out of equilibrium, and vice versa. 
The equilibrium condition thus translates to the following constraints on the coupling $f_N$:
 \begin{eqnarray}
 &(i)& f_N > \frac{M_\Phi}{M_P}~~({\rm in~ equilibrium}), \\\nonumber 
 &(ii)& f_N < \frac{M_\Phi}{M_P}~~({\rm out~ of~ equilibrium}) .
  \end{eqnarray}

Lastly, there are two distinct cases depending on the relative masses of $\Phi$ and $N_{2,3}$:
\begin{eqnarray}
&(a)&  M_N \sim M_\Phi/2~~~( {\rm produced}~N~{\rm is~ non-relativistic}), \\\nonumber
&(b)&  M_N \ll M_\phi  ~~~( {\rm produced}~N~{\rm is~ highly~ boosted}) .
\end{eqnarray}
There are, in total, eight possibilities for matter-antimatter asymmetry and dark matter relic density generation in our model.  
In case (2) with case $(a)$, the $N_{2,3}$ decay immediately after $\Phi$ decay produces $N_{2,3}$, 
for both cases $(i)$ and $(ii)$. 
Since the latter decay to SM fields which have gauge interactions, thermal equilibrium is established immediately, leading to the start of the thermal expansion phase of the universe at the decay epoch of $\Phi$. The treatment of Ref.~\cite{stubbs, MO1} can then be taken over to discuss the baryon asymmetry of the universe. 
This discussion, therefore, covers two of the eight cases. In some sense, this is the simplest case. 
We, therefore, discuss it in the main body of the paper as an illustration of how our scenario works, relegating the remaining cases to the appendix. 
Note that as long as $N_{2,3}$ decay immediately after their production, our results in the main body of the paper remain the same even for  case $(b)$.

\section{Affleck-Dine leptogenesis in case (2)}
In this section, we discuss case (2) above combined with case $(a)$. As just noted, the sub-cases $(i)$ and $(ii)$ are also covered under this scenario. Before getting into that,  we briefly review the main features of AD leptogenesis. In the inflationary phase of the universe,  the real and imaginary parts of the field $\Phi$ start out with non-zero values caused by either primordial Planck scale fluctuations or some dynamical mechanism related to inflaton coupling~\cite{gorbunov}. This fulfills the criterion of CP violation in the Sakharov prescription for baryo- or leptogenesis. In the presence of the lepton number breaking mass term of $\Phi$, this leads to a final asymmetry between the $\Phi$ and $\bar{\Phi}$ fields, which translates into the lepton symmetry during the epoch when the $\Phi$ field oscillates. As $\Phi$ field decays to $N$, this lepton number translates to an asymmetric abundance between $N$ and $\bar{N}$ fields. This further translates to the familiar standard model lepton asymmetry when $N_{2,3}$ decay to $\ell+\chi_L$. This decay takes place above the SM sphaleron decoupling temperature when the sphalerons convert lepton asymmetry to baryon asymmetry of the universe~\cite{Misha, fuku}.

To see this in detail, we start with the lepton asymmetry at $t=\tau_{\Phi}$ given by \cite{stubbs, MO1}
\begin{eqnarray}
 N_L(\tau_\Phi) = 4 Q_L (\epsilon M_\Phi) \frac{|\Phi_I|^3}{|\Phi_*|}\frac{\Gamma_\Phi}{8 \epsilon^2 M_\Phi} {\rm sin 2} \theta, 
\end{eqnarray}
where $Q_L$ is the leptonic charge of $\Phi$, 
$\Phi_I$ is the inflation value when it starts oscillating after the end of inflation, 
$\Phi_*=  M_\Phi/\sqrt{\lambda_\Phi}$ when the quadratic term starts dominating the potential, 
and $\tan\theta=\frac{{\rm Im}[\Phi_I]}{{\rm Re}[\Phi_I]}$. Using this expression, we can evaluate the co-moving lepton asymmetry as
\begin{eqnarray}
n_L(\tau_\Phi)\simeq N_L\left(\frac{a_I}{a(\tau_\Phi)}\right)^3=
N_L\left(\frac{a_I}{a_*}\right)^3 \left(\frac{a_*}{a_(\tau_\Phi)}\right)^3=
N_L\left(\frac{\Phi_*}{\Phi_I}\right)^3 \left(\frac{H(\tau_\Phi)}{H_*}\right)^2\simeq 3Q_L \frac{\Gamma^3_\Phi M^2_P}{\epsilon M^2_\Phi},
\end{eqnarray}
where we have used $H(\tau_\Phi)=\Gamma_\Phi$ and $(H_*)^2=\frac{M^2_\Phi \Phi^2_*}{3M^2_P}$. 
Since the universe gets thermalized at $t \sim \tau_\Phi$, 
the entropy density at this time is given by $s(\tau_\Phi) \sim \frac{4}{3} \rho_\Phi(\tau_\Phi)/T_R$, 
where $\rho_\Phi$ is the inflation energy density, and $T_R$ is the reheating temperature, 
which is roughly given by $T_R \sim \sqrt{\Gamma_\Phi M_P}$. 
Thus, we find the lepton asymmetry of the universe at the epoch above the electroweak phase transition as ~\cite{stubbs,MO1}
\begin{eqnarray}
\frac{n_L}{s}~\sim ~\frac{T^3_R}{\epsilon M^2_\Phi M_P}. 
\end{eqnarray}

We also now calculate the reheat temperature $T_R$ in terms of the parameters of the model. Since as soon as $\Phi$ decays, the universe thermalizes, and $T_R$ is given by
\begin{eqnarray}
\rho_{rad}\simeq \rho_\Phi \sim 3\Gamma^2_\Phi M^2_P, 
\end{eqnarray}
leading to 
\begin{eqnarray}
T_R\simeq \left(\frac{90}{g_*}\right)^{1/4}\frac{f_N}{2\pi}\sqrt{\frac{M_P}{M_\Phi}} M_\Phi\equiv KM_\Phi. 
\end{eqnarray}
Clearly, $K\equiv T_R/M_\Phi < 1$ to avoid washout of the lepton asymmetry generated by $\Phi$ decay. 
We can now write
\begin{eqnarray}
\frac{n_B}{s}\simeq \frac{T^3_R}{\epsilon M^2_\Phi M_P} = \frac{K^3M_\Phi}{\epsilon M_P}\sim 10^{-10},
\end{eqnarray}
to reproduce the observed baryon asymmetry of the universe. 
This implies that
\begin{eqnarray}
\frac{M_\Phi}{M_P}\simeq \frac{\epsilon}{K^3}\times 10^{-10} 
\end{eqnarray}
for $\epsilon \ll 1$ and $K < 1$. 

\subsection{Dark matter relic density generation} 
In this subsection, we discuss the generation of relic density and the constraints implied by it on the parameters of the model. The basic procedure is that the $\Phi$ field decays to $N_1N_1$ pair along with the heavier singlet fermions. Since the $N_1$ fields have no  coupling with any of the standard model fields, they simply ``hang around" with their density decreasing with the expansion of the universe. The DM particles are not in equilibrium. To get the relic density, we first note that after $\Phi$ decay, 
we have $n_{DM}(\tau_\Phi)= 2 \, {\rm Br}(\Phi\to N_1N_1) \, n_\Phi(\tau_\Phi)$. 
We estimate the number density of inflaton as
\begin{eqnarray}
n_\Phi (\tau_\Phi)=\frac{\rho_\Phi(\tau_\Phi)}{M_\Phi}\simeq \frac{\rho_{rad}(\tau_\Phi)}{M_\Phi}.
\end{eqnarray}
Using $s(\tau_\Phi)\simeq \frac{4}{3}\rho_{rad}(\tau_\Phi)/T_R$, we get
\begin{eqnarray}
Y_{DM}(\tau_\Phi)=\frac{n_{DM}(\tau_\Phi)}{s(\tau_\Phi)}
=\frac{3}{2}\frac{T_R}{M_\Phi} {\rm Br}(\Phi\to N_1N_1). 
\label{Eq:Yield}
\end{eqnarray}
We now express the observed value of $\Omega_{DM}h^2\simeq 0.12$ by using the formula
\begin{eqnarray}
\Omega_{DM}h^2\simeq \frac{s_0 M_{DM} Y_{DM}(\tau_\Phi)}{\rho_c/h^2},
\end{eqnarray}
where $M_{DM}$ is the DM mass, $s_0=2890/{\rm cm}^3$ is the entropy density of the present universe, 
and $\rho_c/h^2 =1.05 \times 10^{-5}$ GeV/cm$^3$ is the critical density. 
This leads to the constraint of 
\begin{eqnarray}
M_{DM}({\rm GeV}) \, K \, {\rm Br}(\Phi\to N_1N_1)\simeq 2.9 \times 10^{-10}. 
\end{eqnarray}

Another constraint can also be derived on the mass of the DM from the considerations of structure formation as follows. To reproduce the large-scale structure in the present universe, the free streaming scale of the DM must satisfy the constraint $\lambda_{FS} \leq 0.1$ Mpc at the epoch of radiation-matter equality \cite{Lin:2000qq}, which also translates to a limit on the DM velocity at the current epoch to be $v_0\leq 10^{-7}$ in units of velocity of light $c=1$. We now translate the velocity limit to DM mass.
In our scenario, DM is highly boosted when it is created by the $\Phi$ decay since it is so much lighter that the inflaton field $\Phi$. We get
\begin{eqnarray}
v_0\simeq \frac{M_\Phi}{2 M_{DM}} \left(\frac{a(\tau_\Phi)}{a(t_0)}\right) = \frac{M_\Phi/2}{M_{DM}} \left(\frac{T_0}{T_R}\right) = \frac{T_0}{2KM_{DM}}, 
\end{eqnarray}
which leads to 
\begin{eqnarray}
M_{DM}({\rm GeV}) \, K \geq 1.2 \times 10^{-6} .
\end{eqnarray}
Combining this with the DM relic density constraint, we find an upper bound on the branching ratio of $\Phi$ to DM pair as
\begin{eqnarray}
{\rm Br}(\Phi\to N_1N_1)\leq 2.5 \times 10^{-4} .
\end{eqnarray}
In our scenario, the DM is a warm dark matter. For the choice of the benchmark set of parameters in Table I, the dark matter mass is 300 keV. Note that the $M_\Phi$ and $M_N$ are within an order of magnitude of each other, with $\Gamma_N > \Gamma_\Phi$. 


\begin{table}
\begin{center}
\begin{tabular}{|c||c|}\hline\hline
Parameter & set \\\hline
$M_\Phi$  & $10^{12}$ GeV\\
$M_{N_{2,3}}$ & $10^{11}$ GeV  \\
$M_{N_1}$ & 300 keV  \\
$v_R$ & 100-1000 TeV \\
$f_N$ & $10^{-5}$ \\
$\epsilon$ & $ 10^{-3.5}$\\ 
$T_{R}$ & $10^{9.5} $ GeV\\\hline
\end{tabular}
\end{center}
\caption{ Benchmark set of parameters for case (2)  to illustrate that the model works.}
\end{table}

\section{Discussion} 
In this section, we comment on some aspects of the model:

\begin{itemize}

\item The actual anomaly-free symmetry of the model is $SU(3)_c \times SU(2)_L\times SU(2)_R\times U(1)\times U(1)_{B-L}$ where we gauge only the first four symmetries. The fact that the last $U(1)_{B-L}$ symmetry 
is anomaly-free allows us to have the neutrinos be Dirac fermions since it will prevent the Planck scale induced $B-L$ breaking terms in the low energy Lagrangian.

\item  One assumption we have made in discussing the $N_1$ as the dark matter candidate is that it remains secluded from the SM as well as its other ``siblings" $N_{2,3}$. One way to enforce that would be to have the symmetry $N_1 \to \gamma_5 N_1$ so that it is massless but as the Planck scale $\gamma_5$ breaking terms are switched on, there can be new terms in the effective low energy theory of the form $(\bar{N}_! N_1 \chi^\dagger_R\chi_R)/M_P$ which can give it a mass of order keV as required. This symmetry will also keep it secluded from the $N_{2,3}$ in the absence of gravity effects. The Planck scale effects will also generate its mixing with $N_{2,3}$ via terms like $(\bar{N}_1 N_2 \chi^\dagger_R\chi_R)/M_P$. These terms can make $N_1$ and $N_{2,3}$ mix with mixing angle of order $10^{-6}$ and make $N_1$ unstable. The $N_1$ develops decay modes to  $3\nu$ and $\nu+\gamma$ resulting from these mixing terms. We estimate the $\nu+\gamma$ decay mode to be much more dominant over the $3\nu$ mode. The $N_1$ has a lifetime of $10^{33}$ sec. for $M_{N_1}$ less than an MeV due to this mode making $N_1$ quite acceptable as an unstable dark matter. The current lower limit on the lifetime of dark matter in the mass range of a few keV to MeV is about $10^{26}$ sec.~\cite{essig}.

\item The model has three extra light neutrinos, the right-handed components of $\nu$ and 
the dark matter $N_1$.  
All of them decouple from the thermal plasma
at the very early epoch of the universe (the dark matter has never been in thermal equilibrium).
This $\Delta N_{eff}\simeq 0.14$~\cite{babu1}  value can be probed by future precision CMB experiments such as CMB-S4 \cite{CMB-S4:2016ple}.

\item In the model, parity symmetry is softly broken by the Higgs mass terms. As a result, it has no domain wall problem.

\item Due to the Dirac nature of the neutrinos, neutrinoless double beta decay is forbidden in the model. However the model breaks lepton number by four units due to the presence of the $L$-breaking term $\epsilon M^2 \Phi\Phi$. This leads to neutrinoless quadrupole beta decay processes where we have
$(N,Z)\to (N, Z+4)+4 e^-$~\cite{Heeck}. The rates for these processes are however highly suppressed due to $M^{14}$ power in the effective operator as well as the small parameter $\epsilon$.

\end{itemize}

\section{Summary}
In summary, we have presented a unified description of inflation, dark matter, and the origin of matter in a model that solves the strong CP problem by using parity but without any need for the axion. 
The model has four new vector-like fermions per generation which are singlets under the electroweak $SU(2)_{L,R}$ groups plus a complex gauge singlet scalar field that helps with generating the origin of matter through the Affleck-Dine mechanism. The model predicts a Dirac neutrino with all associated consequences for cosmology and colliders.

\section*{Acknowledgement}
This work is supported in part by the United States Department of Energy Grants DE-SC0016013 (K.S.B) and DE-SC0012447 (N.O).


\section*{Appendix}

In this appendix, we analyze the remaining cases. 
Of these, there are two primary cases to be discussed here belonging to the situation when the $N_{2,3}$ are in equilibrium (case $(i)$) after they appear in $\Phi$ decay and the scenario where they are out of equilibrium (case $(ii)$). The latter case is simpler to deal with, and we focus on it. 
As discussed in the main body of the paper, the dark matter particle $N_1$ has never been in thermal equilibrium. 

We are considering the case with $f_N < M_\Phi/M_P$ so that $N_{2,3}$ are out of equilibrium 
through the $\Phi$ mediated process. 
Before starting our analysis, we first consider a theoretical consistency that $N_{2,3}$ are 
out of equilibrium even through their Yukawa interactions of $y_\nu \bar{\ell} \chi N$.
Using the Dirac seesaw formula, $M_\nu \sim 5 \times 10^{-11}$ GeV and setting $v_R=10^6$ GeV, 
we find $y_\nu^2 \sim M_N/M_P \ll 1$. 
The out-of-equilibrium condition for the decay/inverse decay process of $N \leftrightarrow \ell \, \chi$
at the temperature $T$ is estimated as 
\begin{eqnarray}
  \Gamma_N \frac{M_N}{T} \sim \frac{y_\nu^2}{4 \pi} M_N \frac{M_N}{T} < H(T) \sim \frac{T^2}{M_P} , 
\end{eqnarray}
which leads to $T > M_N$. 
Although one may expect that $N_{2,3}$ can be in thermal equilibrium for $T < M_N$, 
the number density of $N_{2,3}$, if they are in thermal equilibrium, is Boltzmann suppressed
for $T < M_N$, and thus the equilibrium condition is never satisfied. 
The out-of-equilibrium condition for the scattering processes, $N {\bar N} \leftrightarrow \ell {\bar \ell}, 
\chi  \chi ^\dagger$, is much weaker than the condition $T > M_N$ since the scattering process is proportional to $y_\nu^4$. 

\subsection*{A1. Delayed reheating in case $(1)+(ii)+(a)/(b)$}
In section 5, we discussed the case when the inflaton decay products immediately thermalize due to instantaneous decay of $N_{2,3}$. This happens when $\Gamma_{N_{2,3}}\geq \Gamma_\Phi$.
Below, we consider the case when the above condition is not satisfied, i.e.~$\Gamma_{N_{2,3}} < \Gamma_\Phi$.

In typical inflation models, where the inflaton decays to SM particles which quickly thermalize, one can calculate the reheat temperature using the decay width of the inflaton. However, if the inflaton decay products do not thermalize immediately but take time to decay, there is delayed reheating after the end of inflation. 
Here we consider such a case. We will find that in this case, the Dirac seesaw is not adequate, and we have to generate the neutrino masses arising from two loop diagram~\cite{babu1}.

The pre-thermal expansion time span from the decay of $\Phi$ to the decay of $N_{2,3}$  can be split into two parts. Let us set the notation to discuss this. Suppose the epoch at which the $\Phi$ decays to $NN$ states is denoted by $t=\tau_\Phi = (\Gamma_\Phi)^{-1}$, the epoch at which $N_{2,3}$ become non-relativistic is denoted by $t=t_m$, and the epoch at which $N_{2,3}$ decay to SM fields is given by $t=\tau_N$. We can split the total time interval as follows: $\tau_\Phi \ll t_m \ll \tau_N$. There are two distinct time spans given by time span (A): $\tau_\Phi < t < t_m$, and time span (B): $t_m < t < \tau_N$, where we have assumed for simplicity that $\tau_{N_1}\simeq \tau_{N_2}$. It is clear that the universe is radiation-dominated in time span (A), whereas the universe is matter-dominated in time span (B). As a result, in time span (A), we have $a(t)\propto t^{1/2}$ and time span (B) $a(t)\propto t^{2/3}$. The time $t_m$ is determined by 
\begin{eqnarray} 
\frac{1}{2} M_\Phi \left(\frac{a(\tau_\Phi)}{a(t_m)}\right) =M_N ,
\end{eqnarray}
from which we find $t_m = \left(\frac{M_\Phi}{2 M_N} \right)^2 \tau_\Phi$. 
Then, $t_m < \tau_N$ implies that
\begin{eqnarray}
\frac{\Gamma_N}{\Gamma_\Phi} < 4 \, \frac{M^2_N}{M^2_\Phi}.
\end{eqnarray}
If neutrino masses are given by Dirac seesaw, we get $\Gamma_N \sim  \frac{M^2_N}{4\pi M_P}$, leading to the inequality
\begin{eqnarray}
f_N > \frac{1}{2}\left(\frac{M_\Phi}{M_P}\right)^{1/2}.
\end{eqnarray}
This condition conflicts with the out-of-equilibrium condition for $N$ i.e.~$f_N < \frac{M_\Phi}{M_P}$. 
Thus we conclude that in this case, we cannot use the Dirac seesaw for neutrino mass generation but rather the two-loop analysis~\cite{babu1}, so that $\Gamma_N$ becomes a free parameter unrelated to the neutrino mass. 

To discuss lepton asymmetry in this case, we start at $t=\tau_\Phi$ where we have \cite{stubbs, MO1}
\begin{eqnarray}
n_L(\tau_\Phi) \sim 3Q_L \frac{\Gamma^3_\Phi M^2_P}{\epsilon M^2_\Phi}. 
\end{eqnarray}
Using this, we can write
\begin{eqnarray}
\frac{n_L(\tau_N)}{s(\tau_N)}=n_L(\tau_\Phi) \left(\frac{a(\tau_\Phi)}{a(\tau_N)}\right)^3\frac{3T_R}{4 \rho_{rad}(\tau_N)}.
\end{eqnarray}
Using $\left(\frac{a(\tau_\Phi)}{a(\tau_N)}\right)^3= \left(\frac{a(\tau_\Phi)}{a(t_m)}\right)^3 
 \left(\frac{a(t_m)}{a(\tau_N)}\right)^3 =\frac{M_\Phi}{2 M_N}\frac{\Gamma_N^2}{\Gamma_\Phi^2}$ and 
 $\rho_{rad}(\tau_N)=3 \Gamma_N^2 M_P^2$, we can rewrite the above equation as
\begin{eqnarray}
\frac{n_L(\tau_N)}{s(\tau_N)} \sim \frac{3}{8} Q_L \frac{\Gamma_\Phi T_R}{\epsilon M_\Phi M_N} .
\end{eqnarray}
We see that due to delayed reheating, this expression is different from the one in ~\cite{stubbs, MO1}.

\begin{eqnarray}
n_{DM}(\tau_N)= n_{DM}(\tau_\Phi)\left(\frac{a(\tau_\Phi)}{a(\tau_N)}\right)^3
=n_{DM}(\tau_\Phi) \frac{M_\Phi}{2 M_N}\frac{\Gamma_N^2}{\Gamma_\Phi^2}. 
\end{eqnarray}
Then, the DM Yield is given by
\begin{eqnarray}
  Y_{DM}(\tau_N) = \frac{n_{DM}(\tau_N)}{s(\tau_N)}  
  = n_{DM}(\tau_\Phi) \frac{M_\Phi}{2 M_N}\frac{\Gamma_N^2}{\Gamma_\Phi^2} \frac{3 T_R}{4 \rho_{rad}(\tau_N)}
=\frac{3}{4}\frac{T_R}{M_N} {\rm Br}(\Phi\to N_1 N_1). 
\end{eqnarray} 
This result is obtained by replacing $M_\Phi$ to $M_N$ in Eq.~(\ref{Eq:Yield}), 
but $T_R$ is determined by $\Gamma_N= H(\tau_N)= \sqrt{\frac{\rho_{rad}(\tau_N)}{3 M_P}}$ 
with $\rho_{rad}=\frac{\pi^2}{30} g_* T_R^4$, where $g_* \sim 100$ is the effective degrees of freedom
of relativistic particles in the thermal plasma.

\subsection*{A2: Delayed reheating with boosted $N$'s from inflaton decay: case $(1)/(2)+(ii)+(b)$} 
This case arises when $M_N \ll M_\Phi$ so that when $\Phi\to NN$, the $N_{2,3}$ are highly boosted. As a result, their lifetime is dilated by the factor $\frac{M_\Phi/2}{M_N}$. They, therefore, remain relativistic till the time of their decay ($t_{decay}$). This is a crucial difference from A1, where we set $t_m < \tau_N$ 
and $N_{2,3}$ decay after they get non-relativistic. 
In the present case, there is a time span between $\tau_\Phi \leq t \leq t_{decay}$, when the $N_{2,3}$ are not in equilibrium but are relativistic. One can then use the fact that $a(t)\propto t^{1/2}$ during the whole span. Let us derive the formula for $n_B/s$ and the dark matter relic density in this case. 
We will see that we can use the Dirac seesaw to determine neutrino mass. 

We start with some preliminaries: The condition for out-of-equilibrium of the $N_{2,3}$ is $f_N < \frac{M_\Phi}{M_P}$. The velocity of the $N$'s is red-shifted due to the expansion of the universe. 
The lifetime gets shorter, and eventually, the $N$'s decay at time $t_{decay}$, 
which is determined by 
\begin{eqnarray}
t_{decay}~=~\tau_N\left(\frac{M_\Phi}{2M_N}\right) \times \left(\frac{\tau_\Phi}{t_{decay}}\right)^{1/2}, 
\end{eqnarray}
so that
\begin{eqnarray}
t_{decay}~=~\tau_N\left(\frac{M^2_\Phi}{4M^2_N}\frac{\Gamma_N}{\Gamma_\Phi}\right)^{1/3} .
\end{eqnarray}
Consistency requires that  
\begin{eqnarray}
&&t_{decay} > \tau_N~\to ~\frac{\Gamma_N}{\Gamma_\Phi} >~4\left(\frac{M_N}{M_\Phi}\right)^2, \\\nonumber
&&t_{decay} > \tau_\Phi~\to ~\frac{\Gamma_N}{\Gamma_\Phi} < ~\left(\frac{M_\Phi}{2M_N}\right). 
\end{eqnarray}
Similarly, 
%
the condition that the $N$'s decay when they are relativistic implies that 
\begin{eqnarray}
t_m ~=~\frac{1}{4}\frac{M^2_\Phi}{M^2_N}\tau_\Phi > t_{decay} 
\to \frac{\Gamma_N}{\Gamma_\Phi} > 4 \left( \frac{M_N}{M_\Phi}\right)^2, 
\end{eqnarray}
which is the same condition derived from $t_{decay} > \tau_N$. 
As long as $\frac{\Gamma_N}{\Gamma_\Phi}$ satisfies the above conditions for parameters of the model, the scenario works. Note that we can consider both cases $(1)$ and $(2)$. 

We now proceed to derive an expression for $n_B (t_{decay})/s(t_{decay})$ as follows:
\begin{eqnarray}
n_B(\tau_\Phi)~=~\frac{3}{2}Q_\Phi {\rm sin}~ 2\theta \frac{\Gamma^3_\Phi M^2_P}{\epsilon M^2_\Phi} .
\end{eqnarray}
Noting that 
\begin{eqnarray}
n_B(t_{decay})~=~n_B(\tau_\Phi)\left(\frac{a(\tau_\Phi)}{a(t_{decay})}\right)^3, 
{\rm and}~~s(t_{decay})=\frac{4}{3}\frac{\rho(\tau_\Phi)}{T_R}\left(\frac{a(\tau_\Phi)}{a(t_{decay})}\right)^4,
\end{eqnarray}
we have
\begin{eqnarray}
\frac{n_B(t_{decay})}{s(t_{decay})}= n_B(\tau_\Phi)\frac{3}{4}\frac{T_R}{\rho(\tau_\Phi)}\left(\frac{a(t_{decay})}{a(\tau_\Phi)}\right).
\end{eqnarray}
By using $\rho(\tau_\Phi)\left(\frac{a(\tau_\Phi)}{a(t_{decay})}\right)^4=\rho (t_{decay})=\frac{\pi^2}{30}g_* T^4_R$, we find that
\begin{eqnarray}
\frac{n_B}{s}\simeq \frac{3}{8}\left(\frac{90}{\pi^2 g_*}\right)^{1/4} \, Q_L \, {\rm sin}2\theta \,\frac{(\Gamma_\Phi M_P)^{3/2}}{\epsilon M^2_\Phi M_P}.
\end{eqnarray}
If we define $\tilde{T}_R \equiv \left(\frac{90}{\pi^2 g_*}\right)^{1/4}\sqrt{\Gamma_\Phi M_P}$, then we can write
\begin{eqnarray}
\frac{n_B}{s}\simeq \frac{3}{8}\left(\frac{90}{\pi^2 g_*}\right)^{1/4} Q_L \,  {\rm sin} 2\theta \, \frac{{\tilde{T}_R}^3}{\epsilon M^2_\Phi M_P} ,
\end{eqnarray}
which is same as the formula in the case $M_N\sim M_\Phi/2$ and $\Gamma_N > \Gamma_\Phi$ once we replace $T_R$ by $\tilde{T}_R$ (see Eq.~(17)).

Similarly, for the dark matter relic density, $Y_{DM}(t_{decay})$, we start with $n_{DM}(\tau_\Phi)$ given by
\begin{eqnarray}
n_{DM}(\tau_\Phi)= 2\frac{\rho(\tau_\Phi)}{M_\Phi} {\rm Br} (\Phi\to N_1N_1)=~6 \frac{\Gamma^2_\Phi M^2_P}{M_\Phi} {\rm Br} (\Phi\to N_1N_1). 
\end{eqnarray}
From this, we get 
\begin{eqnarray}
Y_{DM}(t_{decay})=n_{DM}(\tau_\Phi)\frac{3T_R}{4\rho(\tau_\Phi)}\left( \frac{a(t_{decay})}{a(\tau_\Phi)}\right)
=~\frac{3}{2}\left(\frac{90}{\pi^2g_*}\right)^{1/4} {\rm Br} (\Phi\to N_1N_1) \frac{\sqrt{\Gamma_\Phi M_P}}{M_\Phi}.
\end{eqnarray}
Now using the definition of $\tilde{T}_R$ as above, we find an analogous formula to Eq.~(23) i.e.
\begin{eqnarray}
Y_{DM}(t_{decay})~=~ \frac{3}{2} {\rm Br} (\Phi\to N_1N_1)\frac{\tilde{T}_R}{M_\Phi}.
\end{eqnarray}
This formula is similar to that in Eq.~(24) except that $T_R$ is replaced by $\tilde{T}_R$.



\begin{thebibliography}{99}

\bibitem{PQ} 
R.~D.~Peccei and H.~R.~Quinn,
``CP Conservation in the Presence of Instantons,''
Phys. Rev. Lett. \textbf{38}, 1440-1443 (1977). 

\bibitem{W} 
S.~Weinberg,
``A New Light Boson?,''
Phys. Rev. Lett. \textbf{40}, 223-226 (1978). 


\bibitem{W1}
F.~Wilczek,
``Problem of Strong  $P$  and  $T$  Invariance in the Presence of Instantons,''
Phys. Rev. Lett. \textbf{40}, 279-282 (1978). 


\bibitem{kim} 
J.~E.~Kim,
``Weak Interaction Singlet and Strong CP Invariance,''
Phys. Rev. Lett. \textbf{43}, 103 (1979).


\bibitem{svz}  
M.~A.~Shifman, A.~I.~Vainshtein and V.~I.~Zakharov,
``Can Confinement Ensure Natural CP Invariance of Strong Interactions?,''
Nucl. Phys. B \textbf{166}, 493-506 (1980). 
 
 
\bibitem{dfs} 
M.~Dine, W.~Fischler and M.~Srednicki,
``A Simple Solution to the Strong CP Problem with a Harmless Axion,''
Phys. Lett. B \textbf{104}, 199-202 (1981). 



\bibitem{z}  
A.~R.~Zhitnitsky,
``On Possible Suppression of the Axion Hadron Interactions. (In Russian),''
Sov. J. Nucl. Phys. \textbf{31}, 260 (1980). 


\bibitem{K}   
M.~Kamionkowski and J.~March-Russell,
``Planck scale physics and the Peccei-Quinn mechanism,''
Phys. Lett. B \textbf{282}, 137-141 (1992). 

\bibitem{holman} 
R.~Holman, S.~D.~H.~Hsu, T.~W.~Kephart, E.~W.~Kolb, R.~Watkins and L.~M.~Widrow,
``Solutions to the strong CP problem in a world with gravity,''
Phys. Lett. B \textbf{282}, 132-136 (1992)
[arXiv:hep-ph/9203206 [hep-ph]].


\bibitem{barr} 
S.~M.~Barr and D.~Seckel,
``Planck scale corrections to axion models,''
Phys. Rev. D \textbf{46}, 539-549 (1992). 


\bibitem{BT} 
M.~A.~B.~Beg and H.~S.~Tsao,
``Strong P, T Noninvariances in a Superweak Theory,''
Phys. Rev. Lett. \textbf{41}, 278 (1978). 


\bibitem{MS1}  
R.~N.~Mohapatra and G.~Senjanovic,
``Natural Suppression of Strong P and T Noninvariance,''
Phys. Lett. B \textbf{79}, 283-286 (1978). 


\bibitem{LR1} 
J.~C.~Pati and A.~Salam,
``Lepton Number as the Fourth Color,''
Phys. Rev. D \textbf{10}, 275-289 (1974)
[erratum: Phys. Rev. D ¥textbf{11}, 703-703 (1975)]. 


\bibitem{LR2} 
R.~N.~Mohapatra and J.~C.~Pati,
``Left-Right Gauge Symmetry and an Isoconjugate Model of CP Violation,''
Phys. Rev. D \textbf{11}, 566-571 (1975); 
%
Phys. Rev. D \textbf{11}, 2558 (1975). 


\bibitem{LR3} 
G.~Senjanovic and R.~N.~Mohapatra,
``Exact Left-Right Symmetry and Spontaneous Violation of Parity,''
Phys. Rev. D \textbf{12}, 1502 (1975). 





\bibitem{MR1} 
R.~N.~Mohapatra and A.~Rasin,
``Simple supersymmetric solution to the strong CP problem,''
Phys. Rev. Lett. \textbf{76}, 3490-3493 (1996)
[arXiv:hep-ph/9511391 [hep-ph]].


\bibitem{MR2} 
R.~N.~Mohapatra and A.~Rasin,
``A Supersymmetric solution to CP problems,''
Phys. Rev. D \textbf{54}, 5835-5844 (1996)
[arXiv:hep-ph/9604445 [hep-ph]].


\bibitem{MR3} 
R.~N.~Mohapatra, A.~Rasin and G.~Senjanovic,
``P, C and strong CP in left-right supersymmetric models,''
Phys. Rev. Lett. \textbf{79}, 4744-4747 (1997)
[arXiv:hep-ph/9707281 [hep-ph]].

\bibitem{BDM} 
K.~S.~Babu, B.~Dutta and R.~N.~Mohapatra,
``Solving the strong CP and the SUSY phase problems with parity symmetry,''
Phys. Rev. D \textbf{65}, 016005 (2002)
[arXiv:hep-ph/0107100 [hep-ph]].


\bibitem{BM} 
K.~S.~Babu and R.~N.~Mohapatra,
``A Solution to the Strong {CP} Problem Without an Axion,''
Phys. Rev. D \textbf{41}, 1286 (1990). 

\bibitem{DW} 
A.~Davidson and K.~C.~Wali,
``Universal Seesaw Mechanism?,''
Phys. Rev. Lett. \textbf{59}, 393 (1987). 



\bibitem{hisano} 
J.~Hisano, T.~Kitahara, N.~Osamura and A.~Yamada,
``Novel loop-diagrammatic approach to QCD \ensuremath{\theta} parameter and application to the left-right model,''
JHEP \textbf{03}, 150 (2023)
[arXiv:2301.13405 [hep-ph]].

\bibitem{Berezhiani:1992pq}
Z.~G.~Berezhiani, R.~N.~Mohapatra and G.~Senjanovic,
``Planck scale physics and solutions to the strong CP problem without axion,''
Phys. Rev. D \textbf{47}, 5565-5570 (1993)
doi:10.1103/PhysRevD.47.5565
[arXiv:hep-ph/9212318 [hep-ph]].

\bibitem{craig} 
N.~Craig, I.~Garcia Garcia, G.~Koszegi and A.~McCune,
``P not PQ,''
JHEP \textbf{09}, 130 (2021)
[arXiv:2012.13416 [hep-ph]].

\bibitem{HH} 
%
L.~J.~Hall and K.~Harigaya,
``Higgs Parity Grand Unification,''
JHEP \textbf{11}, 033 (2019)
[arXiv:1905.12722 [hep-ph]]. 




\bibitem{DHH} 
D.~Dunsky, L.~J.~Hall and K.~Harigaya,
``Sterile Neutrino Dark Matter and Leptogenesis in Left-Right Higgs Parity,''
JHEP \textbf{01}, 125 (2021)
[arXiv:2007.12711 [hep-ph]].

\bibitem{dcruz} 
R.~Dcruz,
``Flavor Physics Constraints on Left-Right Symmetric Models with Universal Seesaw,''
[arXiv:2301.10786 [hep-ph]].

\bibitem{Babu:2022jbn}
K.~S.~Babu and R.~Dcruz,
``Resolving ${\bf W}$ Boson Mass Shift and CKM Unitarity Violation in Left-Right Symmetric Models with Universal Seesaw,''
[arXiv:2212.09697 [hep-ph]].

\bibitem{hariw}  
K.~Harigaya and I.~R.~Wang,
``Baryogenesis in a Parity Solution to the Strong CP Problem,''
[arXiv:2210.16207 [hep-ph]].

\bibitem{Babu:1988yq}
K.~S.~Babu and X.~G.~He,
Mod. Phys. Lett. A \textbf{4}, 61 (1989)
doi:10.1142/S0217732389000095

\bibitem{babu1} 
K.~S.~Babu, X.~G.~He, M.~Su and A.~Thapa,
``Naturally light Dirac and pseudo-Dirac neutrinos from left-right symmetry,''
JHEP \textbf{08}, 140 (2022)
[arXiv:2205.09127 [hep-ph]].




\bibitem{yongchao}  
R.~N.~Mohapatra and Y.~Zhang,
``TeV Scale Universal Seesaw, Vacuum Stability and Heavy Higgs,''
JHEP \textbf{06}, 072 (2014)
[arXiv:1401.6701 [hep-ph]].






\bibitem{dirac1} 
M.~Roncadelli and D.~Wyler,
``Naturally Light Dirac Neutrinos in Gauge Theories,''
Phys. Lett. B \textbf{133}, 325-329 (1983). 

\bibitem{dirac2} 
P.~Roy and O.~U.~Shanker,
``Observable Neutrino Dirac Mass and Supergrand Unification,''
Phys. Rev. Lett. \textbf{52}, 713-716 (1984)
[erratum: Phys. Rev. Lett. \textbf{52}, 2190 (1984)]. 


\bibitem{dirac3} 
P.~H.~Gu and H.~J.~He,
``Neutrino Mass and Baryon Asymmetry from Dirac Seesaw,''
JCAP \textbf{12}, 010 (2006)
[arXiv:hep-ph/0610275 [hep-ph]].


\bibitem{dirac4} 
E.~Ma and R.~Srivastava,
``Dirac or inverse seesaw neutrino masses with $B-L$ gauge symmetry and $S_3$ flavor symmetry,''
Phys. Lett. B \textbf{741}, 217-222 (2015)
[arXiv:1411.5042 [hep-ph]].


\bibitem{dirac41} 
E.~Ma and O.~Popov,
``Pathways to Naturally Small Dirac Neutrino Masses,''
Phys. Lett. B \textbf{764}, 142-144 (2017)
[arXiv:1609.02538 [hep-ph]].


\bibitem{dirac43} 
C.~Bonilla and J.~W.~F.~Valle,
``Naturally light neutrinos in $Diracon$ model,''
Phys. Lett. B \textbf{762}, 162-165 (2016)
[arXiv:1605.08362 [hep-ph]].


\bibitem{dirac42} 
D.~Borah and B.~Karmakar,
``$A_4$ flavour model for Dirac neutrinos: Type I and inverse seesaw,''
Phys. Lett. B \textbf{780}, 461-470 (2018)
[arXiv:1712.06407 [hep-ph]].


\bibitem{dirac5} 
E.~Peinado, M.~Reig, R.~Srivastava and J.~W.~F.~Valle,
``Dirac neutrinos from Peccei\textendash{}Quinn symmetry: A fresh look at the axion,''
Mod. Phys. Lett. A \textbf{35}, no.21, 2050176 (2020)
[arXiv:1910.02961 [hep-ph]].


\bibitem{dirac6} 
S.~Jana, P.~K.~Vishnu and S.~Saad,
``Minimal dirac neutrino mass models from $\hbox {U}(1)_{\mathrm{R}}$ gauge symmetry and left\textendash{}right asymmetry at colliders,''
Eur. Phys. J. C \textbf{79}, no.11, 916 (2019)
[arXiv:1904.07407 [hep-ph]].


\bibitem{dirac7} 
Z.~K.~Silagadze,
``Neutrino mass and the mirror universe,''
Phys. Atom. Nucl. \textbf{60}, 272-275 (1997)
[arXiv:hep-ph/9503481 [hep-ph]].

\bibitem{AD} 
I.~Affleck and M.~Dine,
``A New Mechanism for Baryogenesis,''
Nucl. Phys. B \textbf{249}, 361-380 (1985). 

\bibitem{berzukov} 
F.~L.~Bezrukov and M.~Shaposhnikov,
``The Standard Model Higgs boson as the inflaton,''
Phys. Lett. B \textbf{659}, 703-706 (2008)
[arXiv:0710.3755 [hep-th]].

\bibitem{Inf1} 
M.~P.~Hertzberg and J.~Karouby,
``Generating the Observed Baryon Asymmetry from the Inflaton Field,''
Phys. Rev. D \textbf{89}, no.6, 063523 (2014)
[arXiv:1309.0010 [hep-ph]].


\bibitem{Inf2} 
J.~M.~Cline, M.~Puel and T.~Toma,
``Affleck-Dine inflation,''
Phys. Rev. D \textbf{101}, no.4, 043014 (2020)
[arXiv:1909.12300 [hep-ph]].

\bibitem{gorbunov}  
E.~Babichev, D.~Gorbunov and S.~Ramazanov,
``Affleck-Dine baryogenesis via mass splitting,''
Phys. Lett. B \textbf{792}, 228-232 (2019)
[arXiv:1809.08108 [astro-ph.CO]].


\bibitem{stubbs}  
A.~Lloyd-Stubbs and J.~McDonald,
``A Minimal Approach to Baryogenesis via Affleck-Dine and Inflaton Mass Terms,''
Phys. Rev. D \textbf{103}, 123514 (2021)
[arXiv:2008.04339 [hep-ph]].


\bibitem{MO1} 
R.~N.~Mohapatra and N.~Okada,
``Affleck-Dine baryogenesis with observable neutron-antineutron oscillation,''
Phys. Rev. D \textbf{104}, no.5, 055030 (2021)
[arXiv:2107.01514 [hep-ph]].









\bibitem{Misha} 
V.~A.~Kuzmin, V.~A.~Rubakov and M.~E.~Shaposhnikov,
``On the Anomalous Electroweak Baryon Number Nonconservation in the Early Universe,''
Phys. Lett. B \textbf{155}, 36 (1985). 

\bibitem{fuku} 
M.~Fukugita and T.~Yanagida,
``Baryogenesis Without Grand Unification,''
Phys. Lett. B \textbf{174}, 45-47 (1986). 


\bibitem{Lin:2000qq}
W.~B.~Lin, D.~H.~Huang, X.~Zhang and R.~H.~Brandenberger,
``Nonthermal production of WIMPs and the subgalactic structure of the universe,''
Phys. Rev. Lett. \textbf{86}, 954 (2001)
[arXiv:astro-ph/0009003 [astro-ph]].

\bibitem{essig} 
R.~Essig, E.~Kuflik, S.~D.~McDermott, T.~Volansky and K.~M.~Zurek,
``Constraining Light Dark Matter with Diffuse X-Ray and Gamma-Ray Observations,''
JHEP \textbf{11}, 193 (2013)
[arXiv:1309.4091 [hep-ph]].  
 

\bibitem{CMB-S4:2016ple}
K.~N.~Abazajian \textit{et al.} [CMB-S4],
``CMB-S4 Science Book, First Edition,''
[arXiv:1610.02743 [astro-ph.CO]].


\bibitem{Heeck} 
J.~Heeck and W.~Rodejohann,
``Neutrinoless Quadruple Beta Decay,''
EPL \textbf{103}, no.3, 32001 (2013)
[arXiv:1306.0580 [hep-ph]].



\end{thebibliography}
\end{document}